\documentclass[twocolumn,         
               showpacs,            
               preprintnumbers,     
               aps,                 
               prd,          	    
               a4paper,             
               superscriptaddress,      
               nofootinbib,         
               tightenlines,        
               floats,floatfix,11pt
               ]{revtex4-2}              
\usepackage{graphicx}  
\usepackage[margin=0.6in]{geometry}
\usepackage{dcolumn}   
\usepackage{bm}  
\usepackage[normalem]{ulem}
\usepackage[utf8]{inputenc}
\usepackage{amsmath,amssymb}
\usepackage{soul}
\usepackage{float}
\usepackage[thinlines]{easytable}
\usepackage{array,booktabs}
\usepackage{lipsum} 
\usepackage{multirow}
\usepackage{placeins} 
\usepackage[colorlinks=true,linkcolor=blue,citecolor=blue]{hyperref}
\usepackage{subcaption}
\usepackage{array,makecell}
\usepackage{cleveref} 
\usepackage{rotating}

\usepackage{caption}
\usepackage{multirow}
\usepackage{booktabs}
\usepackage{xcolor}
\usepackage{graphicx} 
\begin{document} 


\title{$\delta$-CDM: A Minimal Deformation of $\Lambda$CDM with Scalar Field Reconstruction}

\author{Phichayoot Baisri}
\email{nandan.roy@mahidol.ac.th (Corresponding Author)} 
\affiliation{NAS, Centre for Theoretical Physics \& Natural Philosophy, Mahidol University,
Nakhonsawan Campus, Phayuha Khiri, Nakhonsawan 60130, Thailand}

\author{Nandan Roy}
\email{nandan.roy@mahidol.ac.th (Corresponding Author)} 
\affiliation{NAS, Centre for Theoretical Physics \& Natural Philosophy, Mahidol University,
Nakhonsawan Campus, Phayuha Khiri, Nakhonsawan 60130, Thailand}

\author{Prasanta Sahoo }
\email{prasantmath123@yahoo.com} 
\affiliation{Midnapore College (Autonomous), Midnapore, West Bengal, India, 721101}

\author{Soumya Chakrabarti}
\email{soumya.chakrabarti@vit.ac.in}
\affiliation{School of Advanced Sciences, Vellore Institute of Technology, \\ 
Tiruvalam Rd, Katpadi, Vellore, Tamil Nadu 632014 \\ India}

\author{Jackson Levi Said}
\email{jackson.said@um.edu.mt}
\affiliation{Institute of Space Sciences and Astronomy, University of Malta, Malta, MSD 2080}
\affiliation{Department of Physics, University of Malta, Malta}

\begin{abstract}

Recent DESI BAO observations provide intriguing hints that dark energy may be dynamical in nature. To investigate deviations of the dark energy equation of state (EoS) from $w = -1$, we introduce the $\delta$-CDM framework, a controlled deformation of $\Lambda$CDM in which deviations from a cosmological constant are parametrized by a redshift-dependent function $\delta(z)$, defined through $w_{\rm de}(z) = -1 + \delta(z)$. As an illustrative example, we reconstruct $\delta(z)$ using effective scalar field dynamics of thawing type, encompassing both quintessence and phantom regimes within a unified description. Notably, the reconstructed $\delta(z)$ is independent of the specific scalar field realization, ensuring theoretical robustness. Using Planck CMB-SPA data, DESI DR2 BAO measurements, and the Pantheon+ supernova sample within a Bayesian Markov Chain Monte Carlo analysis, we find that the $\tilde{w}_0\tilde{w}_a$ parametrization is preferred over this thawing-type realization of deviations from $w = -1$. Overall, the $\delta$-CDM framework provides a minimal yet flexible extension of $\Lambda$CDM, capable of capturing late-time dynamical features of dark energy.
\end{abstract}
\maketitle

\section{Introduction}

The discovery of the late-time accelerated expansion of the Universe through observations of Type Ia supernovae \cite{SupernovaSearchTeam:1998fmf,SupernovaCosmologyProject:1998vns} marked a profound turning point in modern cosmology. Within the framework of general relativity, this acceleration is attributed to dark energy, which today comprises nearly 70\% of the total cosmic energy budget \cite{Planck2020}. The concordance cosmological model, $\Lambda$CDM, accounts for this phenomenon through a cosmological constant $\Lambda$, characterized by a constant equation of state (EoS) parameter $w=-1$. Remarkably, $\Lambda$CDM provides an excellent fit to a wide range of cosmological observations spanning the cosmic microwave background, large-scale structure, and supernova distance measurements. 

Despite its empirical success, $\Lambda$CDM faces persistent theoretical and observational challenges \cite{RevModPhys.61.1,Frieman:2008sn,Carroll:2000fy,Padmanabhan:2002ji,Copeland:2006wr}. From a theoretical perspective, the cosmological constant problem—namely the enormous discrepancy between the observed value of $\Lambda$ and quantum vacuum expectations—and the coincidence problem remain unresolved. On the observational front, discrepancies such as the $H_0$ tension \cite{Riess2019,Verde2019} and the $S_8$ tension \cite{Heymans2021} may indicate that late-time cosmic acceleration is not fully captured by a strictly constant dark energy component.

More recently, baryon acoustic oscillation measurements from DESI have hinted at a possible dynamical evolution of dark energy \cite{DESI:2024mwx,DESI:2025zgx,DESI:2024aqx,Turyshev:2026ewm}, including mild evidence for a crossing of the phantom divide ($w=-1$) in the recent cosmological past citations \cite{Calderon:2024DESIcrossing,Roy:2025DynamicalDE,Berghaus:2024ScalarFieldDESI,Ye:2024PhantomCrossing}. If substantiated, such behavior would be incompatible with a pure cosmological constant and would pose a serious challenge to minimally coupled single canonical scalar field (quintessence) models, which cannot smoothly cross the phantom barrier without introducing pathologies or additional degrees of freedom. These developments therefore motivate a systematic exploration of dynamical dark energy scenarios capable of accommodating such transitions within a theoretically controlled framework.

A wide variety of parametrizations of the dark energy EoS $w(a)$ have been proposed, including the $w_0w_a$ model \cite{Chevallier2001,Linder2003}, the Jassal–Bagla–Padmanabhan (JBP) form \cite{Jassal2005}, logarithmic and transition-based parametrizations~\cite{Efstathiou1999,Hannestad2004}, bounded parametrizations~\cite{Barboza2008}, quintom-inspired descriptions~\cite{Feng2005}, and non-parametric reconstruction approaches~\cite{Zhao2012}. While these parametrizations are phenomenologically useful and flexible, they are often introduced at the level of the background expansion history without an explicit connection to an underlying scalar field dynamics. As a result, the physical interpretation of their parameter space and the theoretical consistency of phantom-crossing behavior can remain obscure.

Several approaches have been developed to reconstruct the dark-energy dynamics directly from observations or from scalar-field theories, including reconstruction of the equation of state, scalar-field potentials, and effective fluid descriptions \cite{Huterer1999,Starobinsky1998,Saini2000,Sahni2006,Shafieloo2006}. In this work, we introduce a new framework, which we call the $\delta$-CDM model. The central idea is to parametrize deviations from the cosmological constant directly through a redshift-dependent deformation function $\delta(z)$ defined by $w_{\rm de}(z) = -1 + \delta(z)$. By construction, $\delta(z)=0$ recovers $\Lambda$CDM exactly, while nonzero values quantify controlled departures from it. This formulation provides a transparent interpretation: rather than prescribing an arbitrary functional form for $w(z)$, the model explicitly measures deviations from the $\Lambda$ limit in a continuous and physically motivated manner.

Going beyond a purely phenomenological description, as an example we reconstruct the effective scalar field dynamics associated with this $\delta$ formalism  for thawing type scalar field dark energy models. Our treatment accommodates both canonical (quintessence) and phantom scalar field models within a unified setup. Importantly, we show that the resulting effective equation of state emerges independently of the specific scalar field realization. 


We confront the $\delta$-CDM model of the thawing scalar field with current observational datasets, including Pantheon+ \cite{Brout:2022vxf}, Pantheon+SH0ES \cite{Brout:2022vxf,Riess:2021jrx,Yuan_2022}, DES Y5 \cite{Abbott2022}, Planck 2018 \cite{Planck2020}, and DESI DR2 BAO measurements \cite{DESI2024} employing a Bayesian Markov Chain Monte Carlo analysis. We assess its statistical performance relative to $\Lambda$CDM using the minimum chi-square statistic and the Akaike Information Criterion (AIC) \cite{Akaike1974,Liddle2007}. In this work our aim is to determine whether controlled deviations encoded through $\delta(z)$ can improve consistency with late-time observations and shed light on the possible dynamical nature of dark energy.  This formalism can also be extended to other dark-energy or modified-gravity scenarios in order to reconstruct effective fluid dynamics of the models.

The paper is structured as follows. In Sec.~\ref{sec:Background_Dynamics}, we formulate the background evolution equations and introduce the dynamical reconstruction framework. The linear perturbation theory for the proposed model is developed in Sec.~\ref{sec:Linear_Perturbation}. Sec.~\ref{sec:Observational_Data_Sets} details the observational data sets and the statistical procedures employed. Our findings and their physical interpretation are reported in Sec.~\ref{sec:Results}, and we close in Sec.~\ref{sec:Conclusion} with a discussion of the wider cosmological significance of the $\delta$-CDM scenario.

\section{Background Dynamics}
\label{sec:Background_Dynamics}
In this section, we outline the theoretical foundation of the proposed $\delta$-deviation formalism, which is built upon a minimally coupled scalar field evolving in a spatially flat Friedmann–Lemaître–Robertson–Walker (FLRW) universe. We derive the associated dynamical equations, recast them in terms of the redshift, and clarify how they relate to the standard cosmological parameters.

\subsection{Background Field Equations}\label{subsec:Background_Field_Equations}

We consider a spatially flat homogeneous and isotropic Universe described by the flat FLRW metric,
\begin{equation}
ds^2 = -dt^2 + a^2(t)\,(dx^2 + dy^2 + dz^2),
\end{equation}
where $a(t)$ is the cosmic scale factor and $t$ denotes cosmic time. The Universe is assumed to be filled with pressureless matter (both dark matter and baryonic matter) and a scalar field $\phi$ that plays the role of dark energy. Scalar-field dark energy models provide a dynamical alternative to the cosmological constant, including inverse power-law quintessence, tracker solutions, and more general evolving equation-of-state scenarios \cite{Ratra1988,Caldwell1998,Zlatev1999,Steinhardt1999}.



For a spatially homogeneous scalar field $\phi=\phi(t)$, the Friedmann equations for the late time Universe become
\begin{subequations}
\begin{align}
H^2 &= \frac{1}{3}\left(\rho_m + \rho_\phi\right), \label{eq:friedmann1} \\
\dot{H} &= -\frac{1}{2}\left(\rho_m + \rho_\phi + p_\phi\right), \label{eq:friedmann2}
\end{align}
\end{subequations}
where $H=\dot{a}/a$ is the Hubble parameter and $\rho_m$ denotes the total matter density. The energy density and pressure of the scalar field are
\begin{align}
\rho_\phi = \frac{\epsilon}{2}\dot{\phi}^2 + V(\phi), \qquad
p_\phi = \frac{\epsilon}{2}\dot{\phi}^2 - V(\phi). \label{eq:rhophipphi}
\end{align}

The scalar field obeys the Klein–Gordon equation,
\begin{equation}
\epsilon \ddot{\phi} + 3\epsilon H \dot{\phi} + \frac{dV}{d\phi} = 0, \label{eq:klein_gordon}
\end{equation}
which follows from energy–momentum conservation, $\nabla_\mu T^{\mu\nu}_{(\phi)} = 0$, where $\mu,\nu = 0,1,2,3$ denote spacetime indices. This equation represents the motion of $\phi$ with a potential $V(\phi)$ under Hubble friction ($3H\dot{\phi}$). Here $\epsilon$ is the switch parameter, for quintessence ($\epsilon = +1$) and for phantom field ($\epsilon = -1$).

The matter energy density evolves as $\rho_m = \rho_{m0}a^{-3}$ for a pressureless fluid, where $\rho_{m0}$ is its present value. Substituting into Eq.~\eqref{eq:friedmann2}, we obtain
\begin{align}
2\dot{H} = -\frac{\rho_{m0}}{a^3} - \epsilon\dot{\phi}^2. \label{eq:dotH}
\end{align}




Using $\dot{H} = \tfrac{1}{2}a\frac{d}{da}(H^2)$, Eq.~\eqref{eq:dotH} can be recast as
\begin{align}
a\frac{d(H^2)}{da} + \frac{\rho_{m0}}{a^3} = -\epsilon\dot{\phi}^2. \label{eq:H2a}
\end{align}
This differential form explicitly links the kinetic term of the scalar field to the derivative of $H^2$ , showing how the field’s motion modifies the cosmic expansion rate. Again the $\dot{\phi}$ can be expressed as
\begin{equation}
\dot{\phi} = aH\,\frac{d\phi}{da}. \label{eq:phia}
\end{equation}
In terms of the redshift $z=1/a-1$, one can write the above equation as
\begin{align}
\frac{d\phi}{dz} = \left[\frac{2E\,dE/dz - 3\Omega_{m0}(1+z)^2}{\epsilon E^2(1+z)}\right]^{1/2}, \label{eq:phiz}
\end{align}
where $E(z)=H(z)/H_0$ and $\Omega_{m0} = \rho_{m0}/(3H_0^2)$.

Equation~\eqref{eq:phiz} can be used to   reconstruct the scalar field ($\phi(z)$) for a given expansion history of the universe through $E(z)$. Also, the potential of the scalar field can be reconstructed by combining the Friedmann equations Eqs.~\eqref{eq:friedmann1} and \eqref{eq:friedmann2} as
\begin{align}
\frac{V(z)}{3H_0^2} &= -\frac{(1+z)}{3}E\frac{dE}{dz} + E^2 - \frac{1}{2}\Omega_{m0}(1+z)^3. \label{eq:Vofz}
\end{align}

The corresponding fractional density parameters can be expressed as below;
\begin{subequations}
\begin{align}
\Omega_m(z) &= \frac{\Omega_{m0}(1+z)^3}{E^2}, \label{eq:Ommz}\\
\Omega_\phi(z) &= 1 - \Omega_m(z), \label{eq:Omphiz}
\end{align}
\end{subequations}
and the dark energy EoS parameter reads
\begin{align}
w_\phi(z) = \frac{\frac{2}{3}(1+z)E\,dE/dz - E^2}{E^2 - \Omega_{m0}(1+z)^3}. \label{eq:wphiz}
\end{align}

In the following we explain how the $\delta(z)$ formulation can be used to reconstruct the scalar field dynamics.

\subsection{Reconstruction of the scalar field dynamics}\label{subsec:Reformulation_in_Scale_Factor}

The central premise of this work is that dark energy exhibits only minimal deviations from a cosmological constant at late time, implemented through its equation-of-state (EoS) parameter. We therefore parametrize the deviation from the $\Lambda$CDM limit as 
\begin{equation}\label{modified_EoS}
    w_{\rm de}(z) = -1 + \delta(z),
\end{equation}

where $\delta(z)$ parametrizes the dynamical deviation from the cosmological constant. We assume $|\delta(z)| < 1$ to ensure that the departure from $w=-1$ remains perturbative and consistent with current observational bounds. Substituting Eq.~\eqref{modified_EoS} into Eq.~\eqref{eq:wphiz} yields the form of the $\delta(z)$
\begin{align}
\delta(z) = \frac{(1+z)}{3\Omega_\phi E^2}\!\left[2E\frac{dE}{dz} - 3\Omega_{m0}(1+z)^2\right]. \label{eq:deltaz}
\end{align}


Physically, $\delta(z)$ determines how the dark energy density evolves with redshift. A positive $\delta(z)$ corresponds to $w>-1$ (quintessence), while a negative $\delta(z)$ yields $w<-1$ (phantom) \cite{Caldwell2002,Caldwell2003}. Phantom dark energy models have attracted considerable attention because they permit super-accelerated expansion and may lead to future singularities or phantom-divide-crossing behavior in more general dark-energy scenarios \cite{Nojiri2005}. In general to allow for a smooth transition between these regimes (``phantom divide crossing’’), $\delta(z)$ must be a continuous, sign-changing function. Here rather than considering a phenomenological form of the $\delta(z)$ we reconstruct the from of it from the scalar field dynamics with thawing nature. Among scalar field models, thawing scenarios are especially compatible with observations because the dark energy equation of state remains close to $w=-1$ at early times and only begins to deviate from this value at late times, thereby leaving the physics of the early Universe largely unaffected \cite{Caldwell2005,Linder2006,Scherrer2008,Chiba2009}.


 One can express the normalized Hubble function, $E(z)$ using the Eq.(\ref{modified_EoS}) as
\begin{align}
E^2(z) = \Omega_{m0}(1+z)^3 + \Omega_{{\rm de},0}
\exp\!\left[3\!\int_0^z \frac{\delta(z')}{1+z'}dz'\right], \label{eq:E2final}
\end{align}
where $\Omega_{{\rm de},0}=1-\Omega_{m0}$. This equation generalizes the $\Lambda$CDM case: for $\delta(z)=0$, it reduces to $E^2(z)=\Omega_{m0}(1+z)^3+\Omega_{\Lambda0}$.

By differentiating Eq.~\eqref{eq:E2final} and inserting the result into Eq.~\eqref{eq:phiz}, we derive the evolution of the scalar field as a function of $\delta(z)$, namely
\begin{align}
\frac{d\phi}{dz} = \frac{\sqrt{3}}{E(1+z)}\!\left[\frac{(E^2-\Omega_{m0} (1+z)^3)\delta(z)}{\epsilon}\right]^{1/2}. \label{eq:dphidz1}
\end{align}

Since we are interested in the thawing behaviour of the dark energy we can consider $\delta \ll 1$ and a slowly varying function and $(E^2-\Omega_{m0} (1+z)^3) \simeq \Omega_{de0}$ at $z<1$. Thus, for low redshift, we can write approximately,
\begin{align}
\frac{d\phi}{dz} \simeq \frac{\sqrt{3}}{E(1+z)}\!\left[\frac{\Omega_{de 0} \delta(z)}{\epsilon}\right]^{1/2}. \label{eq:dphidz2}
\end{align}



The scalar field potential can be reconstructed by differentiating $E^2$ and using Eq.~\eqref{eq:Vofz}, reads
\begin{align}
V(z)= 3H_0^2  (E^2-\Omega_{m0} (1+z)^3) (1 - \frac{1}{2} \delta(z)). \label{eq:Vfinal}
\end{align}

By applying a similar late-time thawing approximation to the potential, we can also express $(E^2 - \Omega_{m0}(1+z)^3) \simeq \Omega_{de0}$, which motivates to adopt the approximate linear parametrization of the potential as
\begin{align}
V(z) = A + B\,\delta(z), \label{eq:VAB}
\end{align}
where $A$ and $B$ are two parameters. Since $V$ carries the dimensions of energy density and $\delta$ is dimensionless, both $A$ and $B$ must likewise have the units of energy density. Within our approximation scheme, $A$ and $B$ seem to be correlated; however, without loss of generality, they can be treated as independent parameters.

\subsection{Effective Parametric Form}\label{subsec:Effective_Parametric_Form}


The Klein–Gordon equation in Eq.(\ref{eq:klein_gordon}), with the parametrized potential given in Eq.(\ref{eq:VAB}) together with Eq.(\ref{eq:dphidz2}) reduces to 
\begin{align}
\frac{d\delta}{dz} - \frac{\tilde{w}_0}{1+z}\,\delta = 0,\label{eq:KG2}
\end{align}
whose general solution is
\begin{align}
\delta(z) = \tilde{w}_{a}(1+z)^{\tilde{w}_0}, 
\end{align}
\begin{align}
    \tilde{w}_0 = 6/(1 + \frac{2B}{3H_0 ^2 \Omega_{de0}}), \label{eq:B_reconstruction}
\end{align}

where  $\tilde{w}_a$ is an integrating constant. From Eq.~\eqref{eq:deltaz}, the dark energy equation of state becomes
\begin{align}
w_{\rm de}(z) = -1 + \tilde{w}_a  (1+z)^{\tilde{w}_0}. \label{eq:EoS}
\end{align}

It is worth emphasizing that, once the thawing behavior of the scalar fields and the $\delta(z)$ formalism are taken into account, the dark energy EoS acquires a power-law type correction. To obtain a thawing-type evolution of the EoS, we must assume $\tilde{w}_0 < 0$, ensuring that for $z \gg 1$ the dark energy equation of state approaches $w_{\rm de}(z) \rightarrow -1$.

An important point is that the Klein–Gordon equation recast in terms of $\delta(z)$, as shown in Eq. (\ref{eq:KG2}), does not involve the switching parameter $\epsilon$. Consequently, the EoS obtained in Eq. (\ref{eq:EoS}) is completely general and is valid for both quintessence and phantom fields. To study the dynamics of this model numerically, one can either solve the approximate Klein–Gordon equation in Eq. (\ref{eq:KG2}) or equivalently use the reconstructed EoS given in Eq. (\ref{eq:EoS}). As shown in \cite{Roy:2022fif}, adopting an effective EoS formulation leads to the same results as solving the full KG equation. Since our goal in this work is to quantify deviations from $\Lambda$CDM, we choose to follow the latter strategy.

\section{Linear Perturbation}\label{sec:Linear_Perturbation}

We consider the linear perturbation in the Newtonian (longitudinal) gauge, following the standard treatment of cosmological perturbation theory developed in Ref.~\cite{MaBertschinger1995}. The effects of dark-energy perturbations on the growth of structure and cosmic microwave background anisotropies have been extensively investigated in the literature \cite{Weller2003,Bean2004}. Scalar perturbations to the Friedmann–Robertson–Walker metric take the form
\begin{equation}
    ds^{2}
    = a(\eta)^{2}\left[-(1 + 2\psi)\,d\eta^{2}
    + (1 - 2\varphi)\,\delta_{ij}\,dx^{i}dx^{j}\right],
    \label{eq:newtonian_metric}
\end{equation}
with $\psi$ and $\varphi$ denoting the Bardeen potentials and $\eta$ the
conformal time.

In the absence of anisotropic stress these potentials coincide,
$\psi = \varphi$.  Throughout this work we assume that the dark energy (DE)
sector does not generate anisotropic stress, although more general modified
gravity scenarios can violate this condition.  In the Newtonian gauge, the
linearized Einstein equations for scalar perturbations may be written as
\begin{equation}
3\mathcal{H}\varphi' + (3\mathcal{H}^{2} + k^{2})\varphi
= -\frac{3\mathcal{H}^{2}}{2}
\left(\Omega_{m}\delta_{m} + \Omega_{\rm de}\delta_{\rm de}\right),
\label{eq:einstein1}
\end{equation}
\begin{equation}
\varphi'' + 3\mathcal{H}\varphi'
+ \left(\frac{2a''}{a} - \mathcal{H}^{2}\right)\varphi
= \frac{3\mathcal{H}^{2}}{2}\,
\frac{\delta p_{\rm de}}{\delta\rho_{\rm de}}\,
\delta_{\rm de},
\label{eq:einstein2}
\end{equation}
where $\mathcal{H} = a'/a$ is the conformal Hubble parameter, primes denote
derivatives with respect to conformal time, and $\delta_{m}$ and
$\delta_{\rm de}$ are the density perturbations of matter and dark energy,
respectively.  On scales well inside the horizon ($k^{2} \gg \mathcal{H}^{2}$)
and during matter domination, Eq.~\eqref{eq:einstein1} reduces to the usual
Poisson equation,
\begin{equation}
k^{2}\varphi \simeq -\frac{3\mathcal{H}^{2}}{2}
\left(\Omega_{m}\delta_{m} + \Omega_{\rm de}\delta_{\rm de}\right).
\label{eq:poisson}
\end{equation}

The evolution of scalar perturbations in a general fluid is governed by the
continuity and Euler equations.  In the Newtonian gauge these read
\begin{equation}
\delta' = -(1+w)\left(\theta - 3\varphi'\right)
         - 3\mathcal{H}\bigg(\frac{\delta p}{\delta\rho} - w\bigg)\delta,
\label{eq:continuity}
\end{equation}
\begin{equation}
\theta' = -\mathcal{H}(1 - 3w)\theta
          - \frac{w'}{1+w}\theta
          + \frac{\delta p / \delta\rho}{1+w}\,k^{2}\delta
          + k^{2}\psi,
\label{eq:euler}
\end{equation}
where $\theta$ denotes the velocity divergence of the fluid.  The
gauge-invariant ratio of pressure to density perturbations is given by
\begin{equation}
\frac{\delta p}{\delta \rho}
= c_{s}^{2}
+ 3\mathcal{H}(1+w)\left(c_{s}^{2} - c_{a}^{2}\right)
\frac{\theta}{k^{2}},
\label{eq:deltap}
\end{equation}
with $c_{s}^{2}$ the physical (rest-frame) sound speed and $c_{a}^{2}$ the
adiabatic sound speed.  Using the background relation $\dot{p} = c_{a}^{2}\dot{\rho}$,
one obtains
\begin{equation}
c_{a}^{2}
= w - \frac{a}{3(1+w)}\frac{dw}{da}.
\label{eq:adiabatic_cs}
\end{equation}

To allow the EoS to cross the phantom divide while maintaining a smooth evolution of perturbations through this transition, we adopt the PPF approximation scheme \cite{Hu:2005Crossing,Fang:2008PPF,Blas:2011rf} as implemented in the \texttt{CLASS} code.

\section{Observational Data Sets}
\label{sec:Observational_Data_Sets}

To evaluate how well the thawing model reconstructed here agrees with current cosmological observations, we carried out an MCMC analysis that simultaneously constrains the standard cosmological parameters and the additional parameters introduced by the model. For this purpose, we implemented the EoS given in Eq.(\ref{eq:EoS}) in the publicly available \texttt{CLASS} code~\cite{Blas:2011rf} \footnote{The code is publically available at \href{https://github.com/Nandancosmos/deltaCDM-DE.git}{https://github.com/Nandancosmos/deltaCDM-DE.git}} and used the cosmological MCMC framework \texttt{Cobaya}~\cite{Torrado:2020dgo} to fit the model to the observational data. The open-source plotting package \texttt{GetDist}  ~\cite{Lewis:2019xzd} was employed to display the resulting posterior distributions. In the following, we describe in detail the dataset adopted in our analysis.

Below, we provide specifics about the dataset used in our analysis.

\subsection{Supernova Data}
Type Ia supernovae are widely used as standard candles because of their comparatively uniform absolute luminosity~\cite{reiss1998supernova,SupernovaSearchTeam:1998fmf}. In the present research, we used the Pantheon Plus compilation of SN-Ia data~\cite{Scolnic:2021amr, Riess:2021jrx, Brout:2022vxf}. The Pantheon+ dataset comprises 1701 light curves with redshifts spanning from $0.01$ to $2.26$. In addition, we also consider the Pantheon+SH0ES compilation, which incorporates the SH0ES Cepheid calibration and provides a direct late-time constraint on the Hubble constant $H_0$~\cite{Riess:2021jrx}.


\subsection{DESI BAO Data}

The density of visible baryonic matter exhibits recurring, periodic fluctuations called baryon acoustic oscillations. These oscillations are essential standard rulers for precise distance measurements in cosmology. In this study, we used the 2025 BAO observation data from the Dark Energy Spectroscopic Instrument (DESI-DR2) as noted in the reference ~\cite{DESI:2025zgx}.  BAO provides measurements of the effective distance along the line of sight as
\begin{equation}
\frac{D_H(z)}{r_d} =\frac{c r_d^{-1}}{H(z)}
\end{equation}

and along the transverse line of sight as,
\begin{equation}
    \frac{D_{M}(z)}{r_{d}}\equiv\frac{c}{r_{d}}\int_{0}^{z}\frac{d\tilde{z}}{H(\tilde{z})}=\frac{c}{H_{0}r_{d}}\int_{0}^{z}\frac{d\tilde{z}}{h(\tilde{z})}.
\end{equation}
The angle average distance is measured as,
\begin{equation}
\frac{D_V(z)}{r_d} =\left[\frac{c z r_d^{-3} d_L^2(z)}{H(z)(1+z)^2}\right]^{\frac{1}{3}}.
\end{equation}

Here, the luminosity distance is denoted by $d_L(z)$. 

\subsection{CMB SPA}

A combined data set that includes the SPT-3G D1 SPTlite likelihood \cite{Balkenhol:2024sbv}, the ACT DR6 full multifrequency likelihood, Planck PR4 data restricted to multipoles $l < 1000$ for temperature and $l < 600$ for polarization, along with the \textit{Plancksroll2EE} likelihood.

\begin{table*}[ht]
\centering
\resizebox{\textwidth}{!}{%

\begin{tabular}{llccccc}
\toprule

Model & Dataset 
& $H_0$ & $\Omega_b h^2$ & $\Omega_c h^2$ & $\Omega_m$ & $\sigma_8$ \\

\midrule

\multirow{5}{*}{$\Lambda$CDM}
& CMB-SPA
& $67.32\pm 0.38$ & $0.022397\pm 0.000095$ & $0.12042\pm 0.00093$ & $0.3166\pm 0.0055$ & $0.8147\pm 0.0039$ \\
& CMB-SPA+PP+DESI
& $68.16\pm 0.24$ & $0.022466\pm 0.000089$ & $0.11836\pm 0.00060$ & $0.3045\pm 0.0033$ & $0.8133\pm 0.0039$ \\
& CMB-SPA+PPS+DESI
& $68.46\pm 0.24$ & $0.022518\pm 0.000093$ & $0.11772\pm 0.00057$ & $0.3006\pm 0.0031$ & $0.8128\pm 0.0038$ \\
& PP+DESI
& $72^{+10}_{-4}$ & $0.0270^{+0.013}_{-0.0046}$ & $0.134^{+0.032}_{-0.020}$ & $0.3047\pm 0.0080$ & $0.79^{+0.16}_{-0.36}$ \\
& PPS+DESI
& $73.8\pm 1.0$ & $0.0285\pm 0.0015$ & $0.1365^{+0.0050}_{-0.0059}$ & $0.3042\pm 0.0078$ & $0.80^{+0.19}_{-0.36}$ \\

\midrule

\multirow{5}{*}{$w_0w_a$}
& CMB-SPA
& $77^{+10}_{-9}$ & $0.022435\pm 0.000097$ & $0.11959\pm 0.00097$ & $0.251^{+0.032}_{-0.085}$ & $0.896^{+0.097}_{-0.064}$ \\
& CMB-SPA+PP+DESI
& $67.71\pm 0.59$ & $0.022429\pm 0.000092$ & $0.11948\pm 0.00078$ & $0.3110\pm 0.0055$ & $0.8157\pm 0.0074$ \\
& CMB-SPA+PPS+DESI
& $69.04\pm 0.54$ & $0.022447\pm 0.000094$ & $0.11964\pm 0.00076$ & $0.2995\pm 0.0049$ & $0.8283\pm 0.0071$ \\
& PP+DESI
& $70^{+10}_{-6}$ & $0.0262^{+0.013}_{-0.0053}$ & $0.125\pm 0.027$ & $0.305^{+0.018}_{-0.012}$ & $0.74^{+0.17}_{-0.35}$ \\
& PPS+DESI
& $73.4\pm 1.0$ & $0.0308^{+0.0024}_{-0.0036}$ & $0.132^{+0.012}_{-0.0090}$ & $0.304^{+0.018}_{-0.013}$ & $0.75^{+0.17}_{-0.35}$ \\

\midrule

\multirow{5}{*}{$\delta$-CDM}
& CMB-SPA
& $74.5^{+5.8}_{-9.7}$ & $0.022404\pm 0.000097$ & $0.12031\pm 0.00098$ & $0.267^{+0.051}_{-0.061}$ & $0.874^{+0.052}_{-0.078}$ \\
& CMB-SPA+PP+DESI
& $67.32\pm 0.58$ & $0.022475\pm 0.000090$ & $0.11813\pm 0.00063 $ & $0.3118\pm 0.0056$ & $0.8052\pm 0.0065$ \\
& CMB-SPA+PPS+DESI
& $68.83^{+0.58}_{-0.66}$ & $0.022514\pm 0.000092$ & $0.11792\pm 0.00066$ & $0.2978\pm 0.0054$ & $0.8171^{+0.0068}_{-0.0080}$ \\
& PP+DESI
& $70^{+10}_{-5}$ & $0.0260^{+0.014}_{-0.0053}$ & $0.125^{+0.030}_{-0.021}$ & $0.3061\pm 0.0083$ & $0.76^{+0.17}_{-0.35}$ \\
& PPS+DESI
& $73.4\pm 1.0$ & $0.0303\pm 0.0017$ & $0.1337\pm 0.0057$ & $0.3058\pm 0.0083$ & $0.77^{+0.16}_{-0.35}$ \\

\midrule

Model & Dataset 
& $w_0/\tilde{w}_0$ & $w_a/\tilde{w}_a$ & $\chi^2_{\min}$ & $\Delta\chi^2_{\min}$ & $\Delta\mathrm{AIC}$ \\

\midrule

\multirow{5}{*}{$\Lambda$CDM}
& CMB-SPA & - & - & $613.5\pm 4.2$ & $0$ & $0$ \\
& CMB-SPA+PP+DESI & - & - & $2036.5\pm 4.6$ & $0$ & $0$ \\
& CMB-SPA+PPS+DESI & - & - & $2116.2\pm 4.9$ & $0$ & $0$ \\
& PP+DESI & - & - & $1418.1\pm 1.9$ & $0$ & $0$ \\
& PPS+DESI & - & - & $1468.4\pm 2.5$ & $0$ & $0$ \\

\midrule

\multirow{5}{*}{$w_0w_a$}
& CMB-SPA & $< -0.596$ & $< -2.31$ & $610.6\pm 4.3$ & $-2.9$ & $1.1$ \\
& CMB-SPA+PP+DESI & $-0.837\pm 0.054$ & $-0.63^{+0.21}_{-0.18}$ & $2028.2\pm 4.7$ & $-8.3$ & $-4.3$ \\
& CMB-SPA+PPS+DESI & $-0.873\pm 0.053$ & $-0.68\pm 0.21$ & $2103.6\pm 4.7$ & $-12.6$ & $-8.6$ \\
& PP+DESI & $-0.882^{+0.055}_{-0.065}$ & $-0.28^{+0.46}_{-0.37}$ & $1415.2\pm 2.7$ & $-2.9$ & $1.1$ \\
& PPS+DESI & $-0.887^{+0.053}_{-0.062}$ & $-0.24^{+0.46}_{-0.34}$ & $1465.2\pm 2.9$ & $-3.2$ & $0.8$ \\

\midrule

\multirow{5}{*}{$\delta$-CDM}
& CMB-SPA & $> -5.89$ & $< -0.272$ & $612.8\pm 4.4$ & $-0.7$ & $3.3$ \\
& CMB-SPA+PP+DESI & $< -5.62$ & $0.116^{+0.065}_{-0.094}$ & $2035.4\pm 4.9$ & $-1.1$ & $2.9$ \\
& CMB-SPA+PPS+DESI & --- & $-0.028^{+0.047}_{-0.061}$ & $2117.1\pm 5.3$ & $0.9$ & $4.9$ \\
& PP+DESI & $< -4.08$ & $0.149^{+0.067}_{-0.097}$ & $1414.9\pm 2.5$ & $-3.2$ & $0.8$ \\
& PPS+DESI & $< -4.13$ & $0.156^{+0.068}_{-0.092}$ & $1464.9\pm 2.8$ & $-3.5$ & $0.5$ \\

\bottomrule
\end{tabular}

}

\caption{Best-fit values and $68\%$ confidence intervals for cosmological parameters obtained from five data combinations.}
\label{tab:bestfit}

\end{table*}


    \begin{figure*}[!hbt]
            \centering
            \includegraphics[width=2\columnwidth]{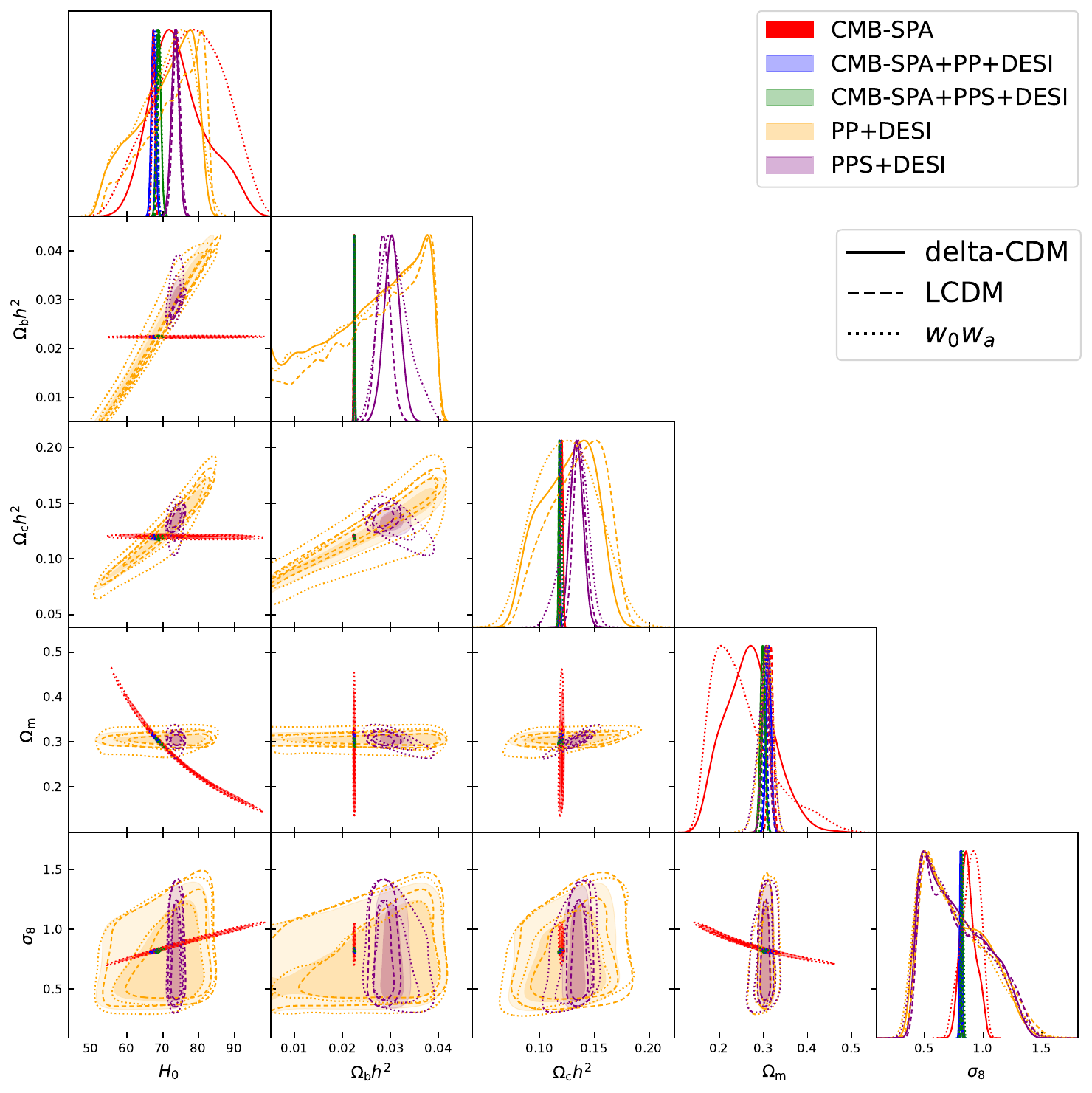}
            \caption{
Observational constraints on $H_0$, $\Omega_b h^2$, and $\Omega_c h^2$ from five data combinations. The contours represent the $68\%$ and $95\%$ confidence regions. Line styles indicate the cosmological models: $\Lambda$CDM (dashed), $\delta$-CDM (solid), and $w_0w_a$ (dotted).
}
            \label{fig:Final_All_Axion}
    \end{figure*}


\begin{figure}[!hbt]
            \centering
            \includegraphics[width=\columnwidth]{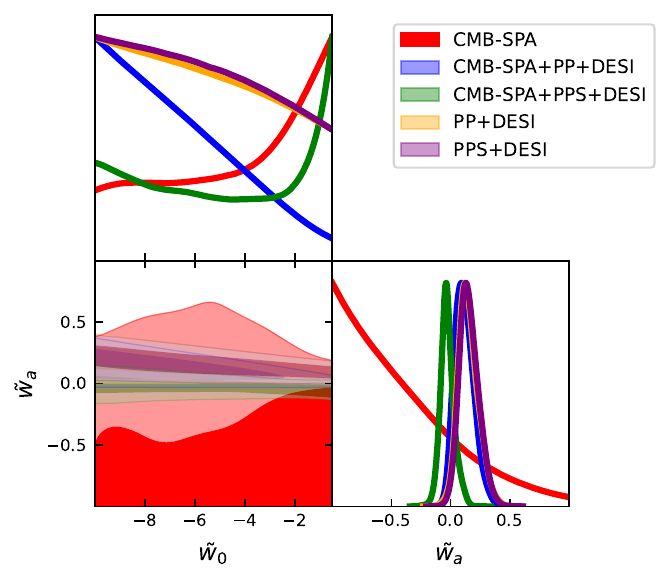}
            \caption{
Triangle plot showing the constraints on the dark energy parameters $\tilde{w}_0$ and $\tilde{w}_a$ in the $\delta$-CDM reconstructed model from five data combinations. The contours correspond to the $68\%$ and $95\%$ confidence regions.
}
            \label{fig:w0wa_constaint}
    \end{figure}


\begin{figure}[!hbt]
            \centering
            \includegraphics[width=\columnwidth]{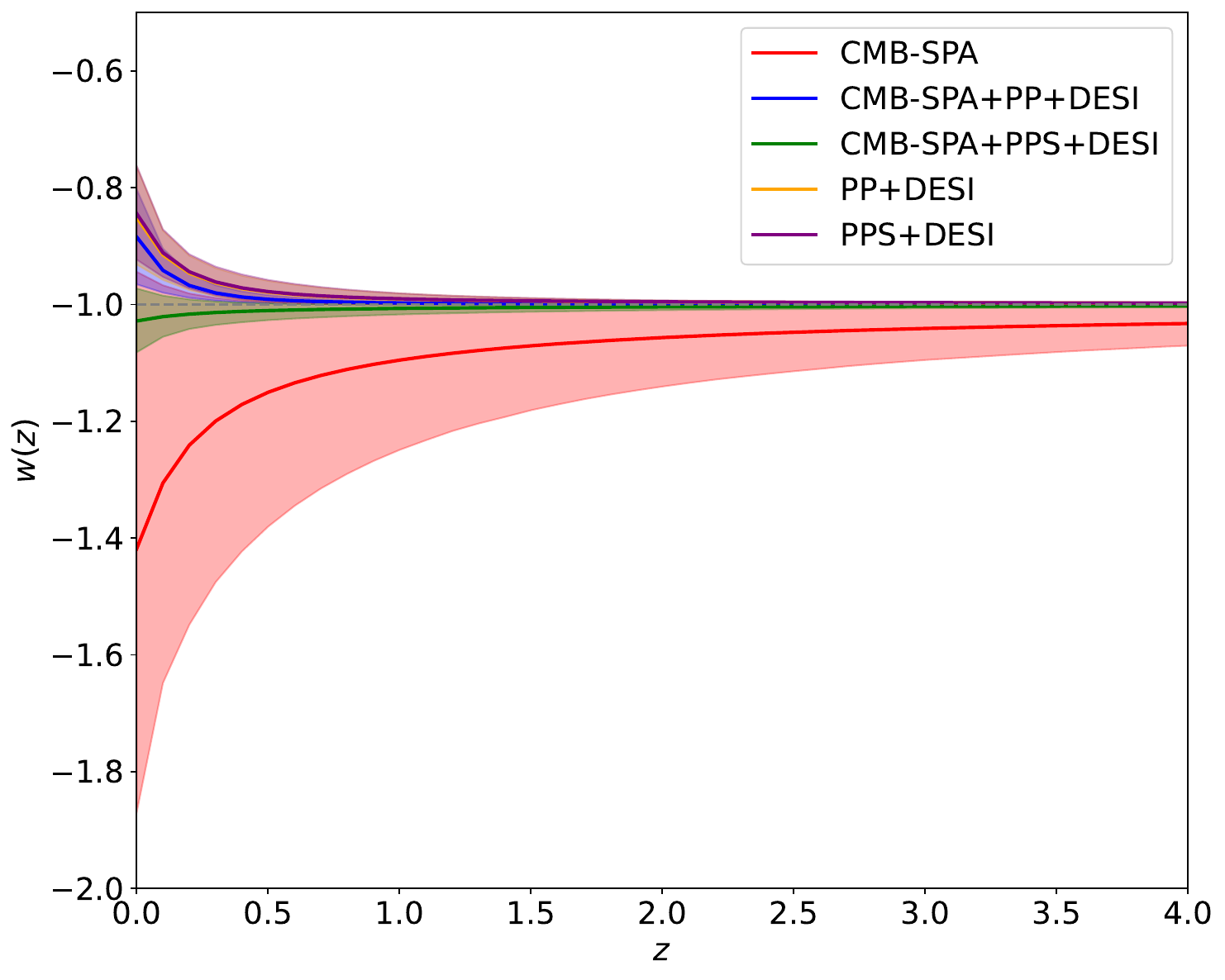}
            \caption{
Evolution of the dark energy equation of state $w(z)$ for the parametrization $w(z)=-1+\tilde{w}_a(1+z)^{\tilde{w}_0}$ from different data combinations. The solid lines represent the mean evolution, and the shaded regions correspond to the $68\%$ confidence intervals.
}
            \label{fig:EoS}
\end{figure}

\begin{figure}[!hbt]
            \centering
            \includegraphics[width=\columnwidth]{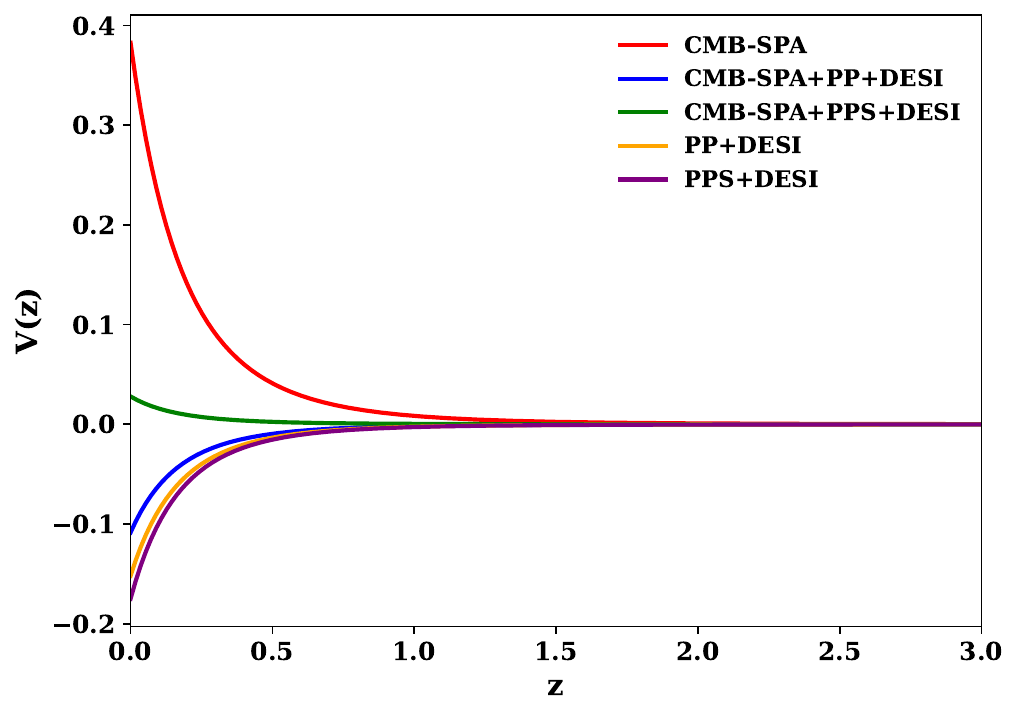}
            \caption{
Evolution of the dimensionless scalar field potential $V(z)=B\delta(z)$, corresponding to Eq. (\ref{eq:VAB}) with A=0, where $\delta(z)=\tilde{w}_a(1+z)^{\tilde{w}_0}$. The background cosmology sets the coefficient $B$ using the Eq.~(\ref{eq:B_reconstruction}) for each dataset, and the solid lines show the corresponding evolution. 
}
            \label{fig:vz}
\end{figure}

\begin{figure}[!hbt]
            \centering
            \includegraphics[width=\columnwidth]{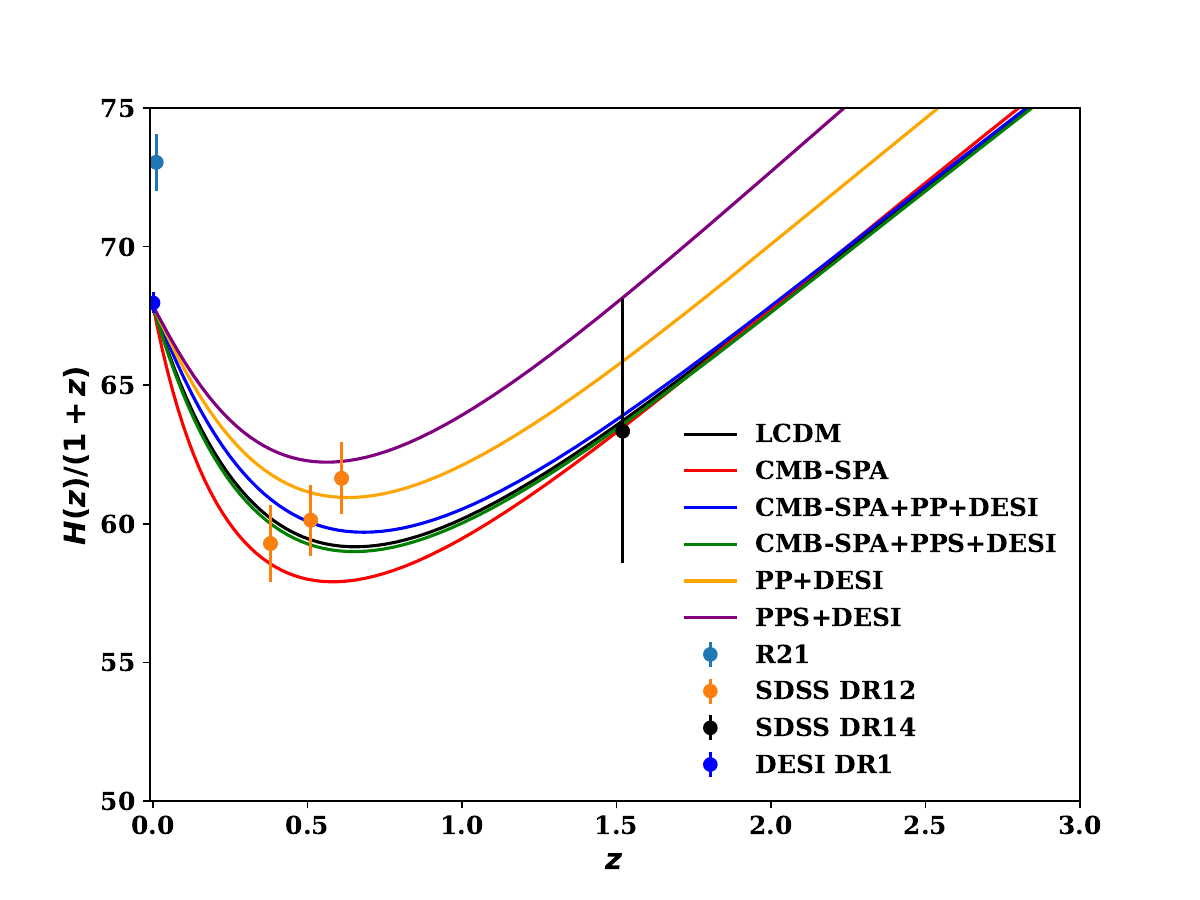}
            \caption{
The Hubble parameter $H(z)$ predicted by the $\delta$-CDM reconstructed model together with observational measurements from R21, SDSS DR12, SDSS DR14, and DESI DR1. The $\Lambda$CDM prediction is also shown.
}
            \label{fig:Hz}
\end{figure}

\begin{figure}[!hbt]
            \centering
            \includegraphics[width=\columnwidth]{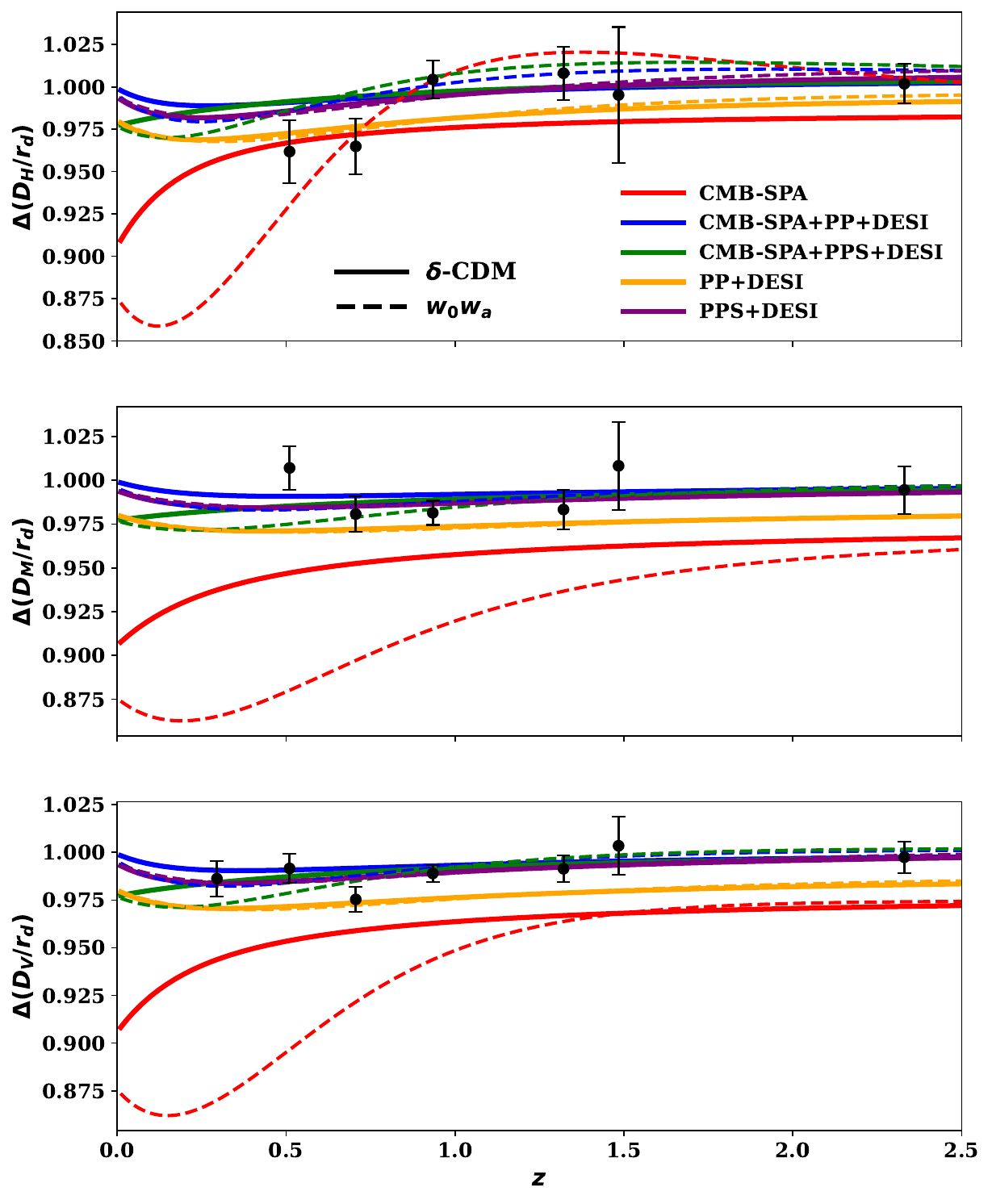}
            \caption{
BAO distance differences of the $\delta$-CDM and $w_0w_a$ model with respect to the $\Lambda$CDM model for different observational constraints. The top panel shows $\Delta(D_H/r_d)$, the middle panel shows $\Delta(D_M/r_d)$, and the bottom panel shows $\Delta(D_V/r_d)$. The observed data from DESI DR2 are shown as black points with corresponding error bars.
}
            \label{fig:BAO}
\end{figure}


\begin{figure}[!hbt]
            \centering
            \includegraphics[width=\columnwidth]{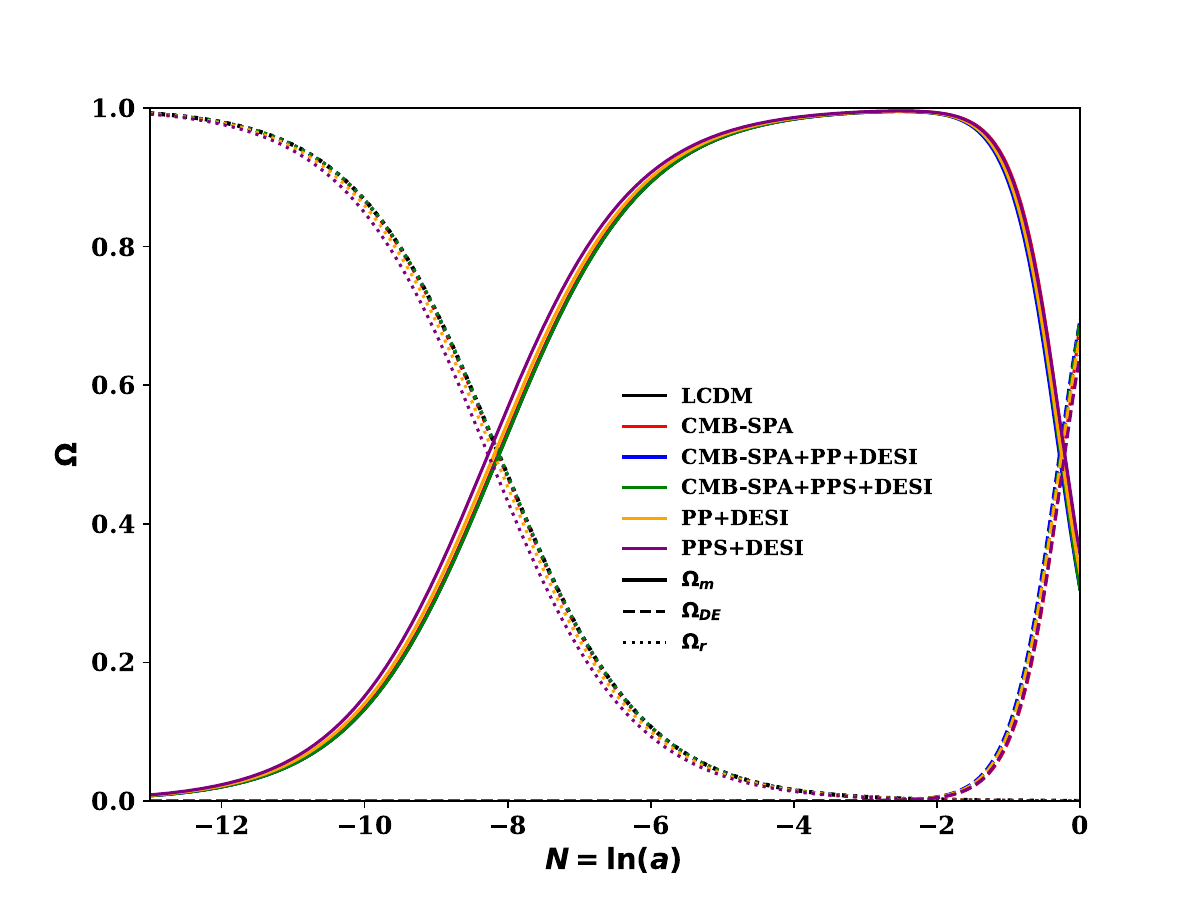}
            \caption{
Evolution of the energy density parameters in the $\delta$-CDM reconstructed model versus $N=\ln(a)$, with the $\Lambda$CDM prediction shown for reference.
}
            \label{fig:density}
\end{figure}

\begin{figure}[!hbt]
            \centering
            \includegraphics[width=\columnwidth]{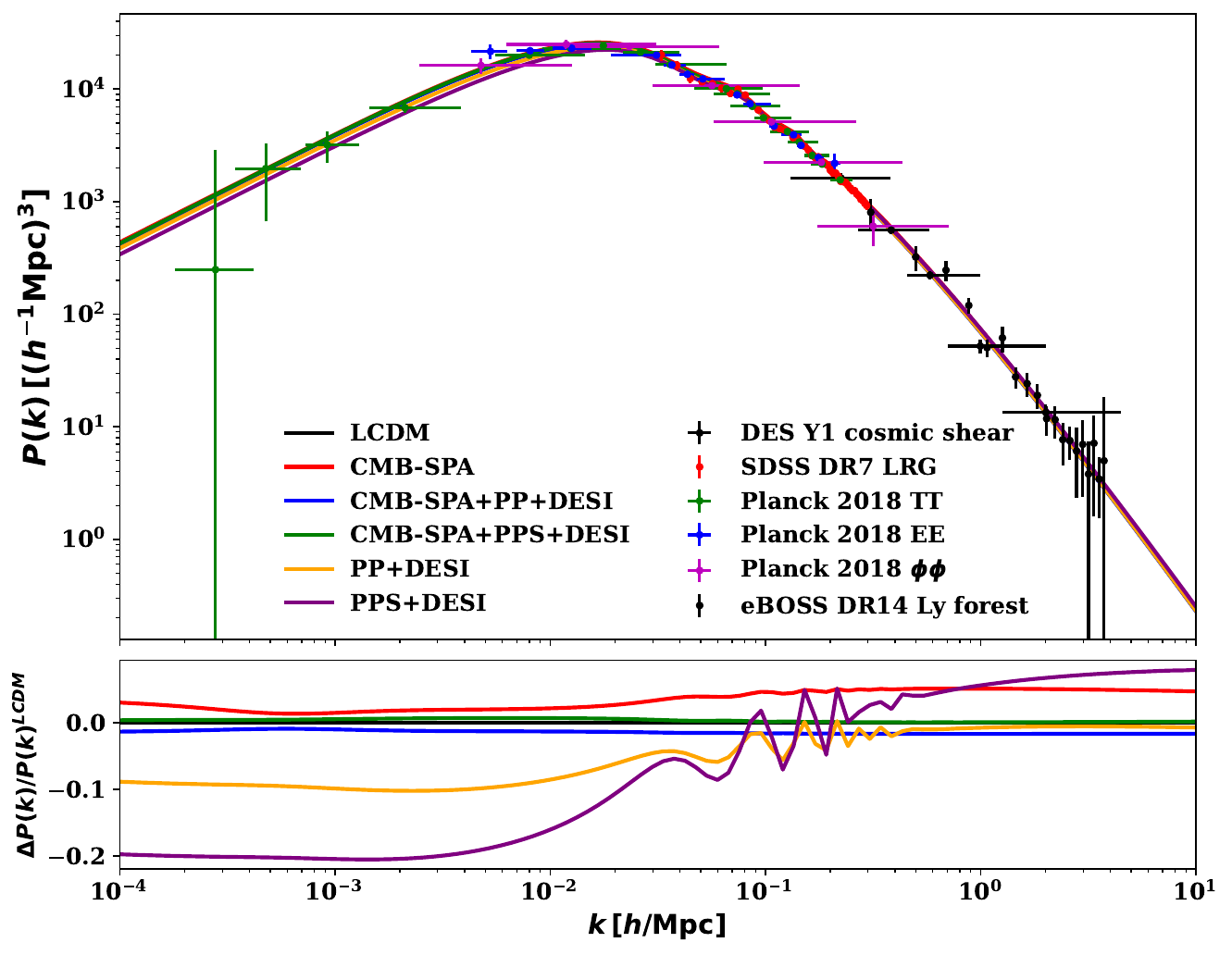}
           \caption{
Matter power spectrum $P(k)$ together with observational measurements. The top panel shows the power spectrum, while the bottom panel presents the relative difference with respect to the $\Lambda$CDM model.
}
            \label{fig:Pk}
\end{figure}

\begin{figure}[!hbt]
            \centering
            \includegraphics[width=\columnwidth]{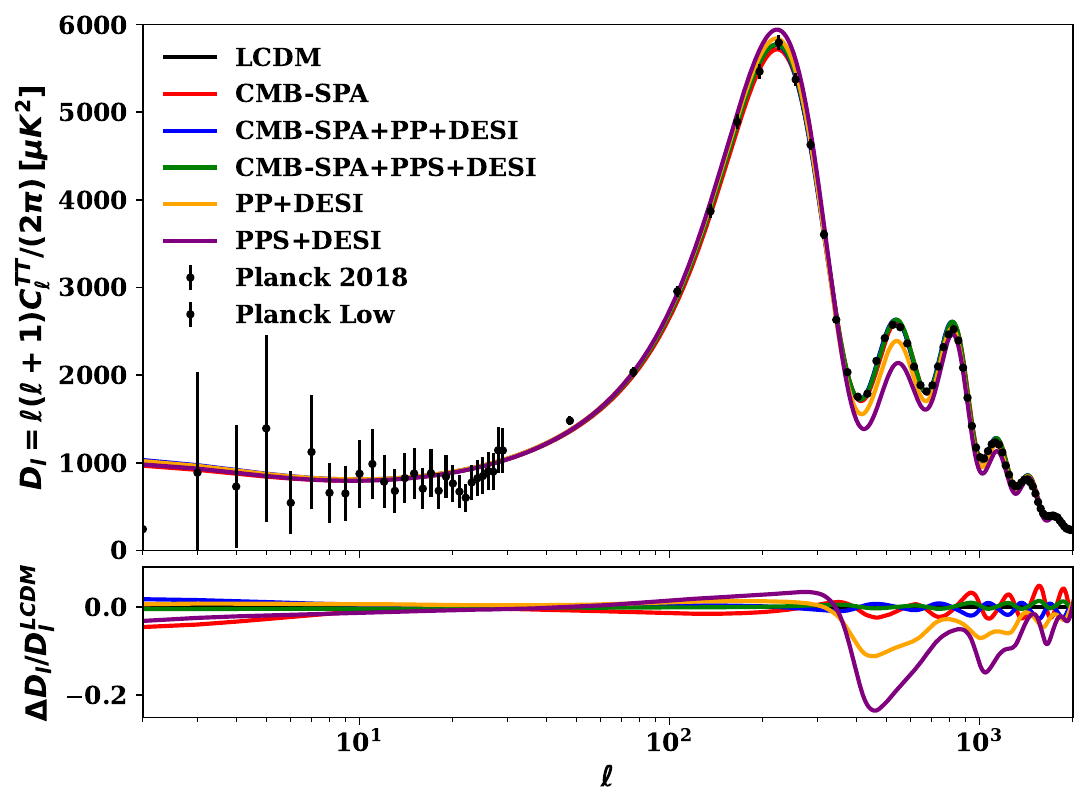}
            \caption{
CMB temperature power spectrum $D_\ell$ together with Planck 2018 measurements. The top panel shows the power spectrum, while the bottom panel presents the relative difference $\Delta D_\ell / D_\ell^{\Lambda\mathrm{CDM}}$.
}
            \label{fig:CMB_TT}
\end{figure}

\section{Results}
\label{sec:Results}
The parameter space of the reconstructed EoS using the $\delta$-CDM formalism is explored using a Markov Chain Monte Carlo (MCMC) analysis implemented in \texttt{\texttt{Cobaya}}, with a modified version of \texttt{CLASS} used to compute the background and perturbation evolution. Our study explored five combinations of cosmological datasets including CMB-SPA, Pantheon Plus (PP), Pantheon Plus SH0ES (PPS) and DESI DR2 (DESI) measurements. To constrain our models, we impose flat priors on the standard cosmological parameters $ H_0 : [40,100], \Omega_{b}h^{2}:[0.005,0.1], \Omega_\mathrm{cdm} h^2:[0.001,0.99],   $ and our choice for the model parameters is $\tilde{w}_0:[-10,-0.5], \tilde{w}_a:[-1, 1]$. The prior on $\tilde{w}_0$ is negative, so that the EoS given in Eq.(\ref{eq:EoS}) does not diverge at higher redshift. The dataset combinations we consider are: CMB-SPA, CMB-SPA+PP+DESI, CMB-SPA+PPS+DESI, PP+DESI, and PPS+DESI.  

The resulting best-fit parameters and their $1\sigma$ confidence intervals are given in Table~\ref{tab:bestfit}. Relative to $\Lambda$CDM, the $\delta$-CDM reconstructed thawing model produces largely compatible constraints on the key cosmological parameters for data sets that include low-redshift observations, while simultaneously accommodating a dynamical dark energy component. From Table~\ref{tab:bestfit}, the CMB-only data show tight constraints on $\Omega_m$ and $H_0$ for $\Lambda$CDM, but for the dynamical dark energy models, $\Omega_m$ and $H_0$ remain unconstrained. Once the low-redshift data are included and $\Omega_m$ and $H_0$ for the dynamical dark energy model are constrained to the value in agreement with the $\Lambda$CDM. Similarly, for the CMB-only data set, the $\sigma_8$ parameter is only weakly constrained in the dynamical dark energy models, whereas adding low-redshift observations renders it compatible with $\Lambda$CDM.

For the $\delta$-CDM reconstructed scenario, the dark energy equation-of-state parameters are constrained to $\tilde{w}_0 < -5.62$ with $\tilde{w}_a = 0.116^{+0.065}_{-0.094}$ for the CMB-SPA+PP+DESI data set, and to $\tilde{w}_a = -0.028^{+0.047}_{-0.061}$ for CMB-SPA+PPS+DESI, while in the latter case $\tilde{w}_0$ remains unconstrained. Regarding the quality of the fit, both dynamical dark energy models, $w_0w_a$ and $\delta$-CDM, provide an improved description of the data relative to $\Lambda$CDM in terms of the $\Delta \chi^2$ statistic. The $w_0w_a$ parametrization typically yields a smaller minimum $\chi^2$ for combinations that include CMB measurements (for instance, $\Delta\chi^2_{\min} = -12.6$ for CMB-SPA+PPS+DESI), whereas for low-redshift-only combinations the $\delta$-CDM reconstruction tends to give slightly lower values (e.g., $\Delta\chi^2_{\min} = -3.2$ for PP+DESI). The only exception occurs for the CMB-SPA+PPS+DESI combination, where for the $\delta$-CDM model the resulting $\Delta \chi^2$ is actually positive, even when compared to $\Lambda$CDM.

However, the corresponding $\Delta \mathrm{AIC}$ values are generally positive, indicating that the current data do not statistically favor the thawing dark energy model reconstructed here using $\delta$-CDM formalism over $\Lambda$CDM despite the mild improvement in $\chi^2_{\min}$. For data combinations without CMB, the $\Delta \mathrm{AIC}$ values are $<1$ for the current model.

Figure~\ref{fig:Final_All_Axion} shows the marginalized posterior distributions for $H_0$, $\Omega_b h^2$, $\Omega_c h^2$, $\Omega_m$ and $\sigma_8$ obtained from the five data combinations. Overall, the constraints from the three cosmological models are broadly consistent. A mild discrepancy appears in the baryon density between the PPS+DESI and CMB-SPA constraints, with $\Omega_b h^2 = 0.0303 \pm 0.0017$ and $\Omega_b h^2 = 0.022404 \pm 0.000097$, respectively. The corresponding difference in the cold dark matter density is relatively smaller, with $\Omega_c h^2 = 0.1337 \pm 0.0057$ for PPS+DESI and $\Omega_c h^2 = 0.12031 \pm 0.00098$ for CMB-SPA. 


Figure~\ref{fig:w0wa_constaint} shows the constraints on the dark energy parameters $\tilde{w}_0$ and $\tilde{w}_a$ for the reconstructed model. The inclusion of DESI DR2 and supernova data significantly tightens the allowed parameter space, reducing the degeneracy between $\tilde{w}_0$ and $\tilde{w}_a$. In the $\delta$-CDM reconstructed model, we obtain $\tilde{w}_0 < -5.62$ with $\tilde{w}_a = 0.116^{+0.065}_{-0.094}$ for the CMB-SPA+PP+DESI combination, and $\tilde{w}_a = -0.028^{+0.047}_{-0.061}$ for CMB-SPA+PPS+DESI and $\tilde{w}_0$ is unconstrained.

The reconstructed evolution of the equation of state is presented in Fig.~\ref{fig:EoS}. In this figure, the $\delta$-CDM reconstruction shows a slight departure from $w=-1$ at late times. As expected, since this model is obtained under the assumption of a thawing single scalar field, no crossing of the phantom divide is observed, even though the functional form of the reconstructed EoS in Eq.~(\ref{eq:wphiz}) in principle allows for such a crossing. This confirms that the $\delta$-CDM reconstruction is consistent with the thawing behavior of a single scalar field.  The CMB-SPA and CMB-SPA+PPS+DESI reconstructions lie in the phantom regime, whereas all other combinations favor quintessence-like behavior. This distinction is also reflected in the reconstructed scalar-field potential shown in Fig.~\ref{fig:vz} considering $A=0$. The CMB-SPA and CMB-SPA+PPS+DESI reconstructions fall within the phantom regime, while all the remaining data combinations support a quintessence-like behavior. In the CMB-SPA and CMB-SPA+PPS+DESI cases, the potential starts from $(V(z)\simeq 0)$ and gradually grows at late times, whereas the CMB-SPA+PP+DESI, PP+DESI, and PPS+DESI reconstructions show a declining potential toward the present epoch. Equation~(\ref{eq:wphiz}) indicates that the EoS depends explicitly on the derivative term $(dE/dz)$, which controls the evolution of the normalized expansion rate $(E(z)=H(z)/H_0)$. 
When the quantity $2E\,\frac{dE}{dz}-3\Omega_{m0}(1+z)^2$ is positive, the scalar field behaves as a quintessence field, while a negative value of $2E\,\frac{dE}{dz}-3\Omega_{m0}(1+z)^2$ corresponds to phantom behavior. From Eq.~(\ref{eq:Vofz}) one sees that, in the quintessence case, once $2E\,\frac{dE}{dz}-3\Omega_{m0}(1+z)^2$ becomes positive, $V(z)$ decreases, whereas in the phantom case, where $2E\,\frac{dE}{dz}-3\Omega_{m0}(1+z)^2$ turns negative at late times, $V(z)$ instead increases.Hence, the reconstructed form of the potential in Eq.~(\ref{eq:VAB}) is consistent with the scalar field dynamics obtained from Eq.~(\ref{eq:Vofz}).

Figure~\ref{fig:Hz} shows the predicted expansion history together with the observational $H(z)$ measurements. The combinations including CMB-SPA closely track the $\Lambda$CDM prediction over the full redshift range. The CMB-SPA+PPS+DESI result lies closest to $\Lambda$CDM, followed by CMB-SPA+PP+DESI and the CMB-SPA-only constraint. These predictions are nearly indistinguishable at high redshift and diverge slightly below $z\approx 1.5$. In contrast, the PP+DESI and PPS+DESI cases remain more clearly separated from the $\Lambda$CDM curve throughout the redshift range considered. 

Figure~\ref{fig:BAO} displays the BAO distance ratios relative to the $\Lambda$CDM prediction. The top, middle, and bottom panels show $\Delta(D_H/r_d)$, $\Delta(D_M/r_d)$, and $\Delta(D_V/r_d)$, respectively. The results for the $w_0,w_a$ model are indicated by the dashed curve. Across all panels, the fit obtained using only the CMB-SPA best-fit parameters provides the poorest agreement with the BAO measurements, whereas the combinations CMB+PP+DESI, CMB+PPS+DESI and PPS+DESI yield a significantly better match to the data.

The background evolution of the energy density parameters is presented in Fig.~\ref{fig:density}. For all parameter choices the sequence of radiation, matter and dark energy domination remains broadly consistent with the $\Lambda$CDM behaviour. The matter power spectrum and the CMB temperature power spectrum are shown in Fig.~\ref{fig:Pk} and Fig.~\ref{fig:CMB_TT}, respectively. The $\delta$-CDM predictions exhibit small deviations from $\Lambda$CDM across all scales while remaining consistent with the observational data. The largest differences in the matter power spectrum occur at low wavenumbers $k \lesssim 10^{-2} h/{\rm Mpc}$ for PP+DESI and PPS+DESI best fit values since the CMB information is not included in these data combinations. For other data set combinations the deviation is significantly small. In Fig.~\ref{fig:CMB_TT}, the CMB temperature anisotropy is shown together with the observed data sets from the Planck mission. The deviations from the $\Lambda$CDM model are presented as fractional differences in the lower panels. These discrepancies are again particularly pronounced for the PP+DESI and PPS+DESI best-fit curves, while for the other data combinations the results remain in close agreement with $\Lambda$CDM.

\section{Conclusion}
\label{sec:Conclusion}

In this work, we introduced the $\delta$-CDM formalism as a framework for reconstructing the effective fluid description of scalar-field dark energy and, more generally, modified gravity scenarios. The key idea is to parameterize deviations from the cosmological constant through a redshift-dependent deformation function $\delta(z)$, defined via the relation $ w_{\rm de}(z)=-1+\delta(z)$. By construction, the $\Lambda$CDM limit is recovered when $\delta(z)=0$, while nonzero values quantify controlled departures from a cosmological constant.

As an illustrative realization of the formalism, we considered thawing scalar-field dynamics and reconstructed the corresponding effective dark-energy equation of state. The reconstruction naturally accommodates both quintessence and phantom scalar fields within a unified framework. An important result is that the reconstructed effective equation of state is independent of the specific scalar-field realization, implying that the resulting parametrization captures the generic late-time dynamics of thawing models rather than the details of any particular scalar-field theory.

Using a Bayesian MCMC analysis, we constrained the model parameters with several combinations of current cosmological observations, including Planck CMB-SPA data, DESI DR2 BAO measurements, and the Pantheon+ and Pantheon+SH0ES supernova compilations. We found that the reconstructed model remains fully consistent with current observations and reproduces the standard cosmological background evolution, matter power spectrum, and CMB anisotropy spectrum with  minimal deviations from $\Lambda$CDM. The inclusion of low-redshift observations with CMB-SPA significantly improves the constraints on the cosmological parameters and reduces degeneracies present in the CMB-only analysis.

From the observational perspective, the reconstructed $\delta$-CDM model yields mild improvements in the minimum $\chi^2$ relative to $\Lambda$CDM for several dataset combinations, particularly when only late-time observations are considered. However, when model complexity is taken into account through the Akaike Information Criterion, the current data do not provide significant statistical evidence in favor of the reconstructed thawing model over the simpler $\Lambda$CDM scenario. Furthermore, a comparison with the $w_0w_a$ parametrization shows that, when CMB observations are included, the $w_0w_a$ model typically offers a better fit to the combined datasets, whereas our current model performs better when only late-time datasets are considered.

The reconstructed evolution of the dark-energy equation of state exhibits only modest departures from $w=-1$, consistent with the thawing nature of the underlying scalar-field dynamics. Depending on the dataset combination, the preferred evolution can lie either in the quintessence or phantom regime, although no phantom-divide crossing is observed in the reconstructed thawing scenario. The corresponding scalar-field potentials display behaviors fully consistent with the reconstructed expansion history and effective equation of state.

Overall, the $\delta$-CDM formalism provides a simple and physically transparent approach for quantifying departures from $\Lambda$CDM by reconstructing the dark-energy dynamics from an underlying physical model rather than introducing purely phenomenological parametrizations. While the thawing realization studied here is not currently favored over $\Lambda$CDM by existing observations, the framework itself remains sufficiently flexible to accommodate a broad class of dark-energy and modified-gravity models. Future high-precision cosmological surveys, including forthcoming DESI, Euclid, Roman Space Telescope, and LSST observations, will further improve constraints on late-time cosmic expansion and may provide a more decisive test of deviations from the cosmological constant \cite{Euclid2019,Spergel2015,LSST2009}. In this context, the $\delta$-CDM framework offers a useful tool for reconstructing and interpreting possible signatures of dynamical dark energy.

\begin{acknowledgements}
JLS would also like to acknowledge funding from ``Xjenza Malta'' as part of the ``Technology Development Programme'' DTP-2024-014 (CosmicLearning) Project.
\end{acknowledgements}

\bibliographystyle{unsrt}
\bibliography{sample}

\end{document}